\documentclass[onecolumn,amsmath,amssymb,amsfonts,a4paper,aps,prd,10pt]{revtex4}
\usepackage{graphicx}
\usepackage{epstopdf}
\usepackage{verbatim}
\newcommand{\nc}{\newcommand}
\nc{\ba}{\begin{eqnarray}} \nc{\ea}{\end{eqnarray}}
\newcommand\be{\begin{equation}}
\newcommand\ee{\end{equation}}
\nc{\D}{\overline{\mbox{D3}}}

\nc{\ga}{\gamma} \nc{\tnu}{\tilde{\nu}} \nc{\tmu}{\tilde{\mu}}

\nc{\x}{{\bf{x}}}

\input{epsf.sty}
\begin{document}

%%%%%%%%%%%%%%%%%%%%%%%%%%%%%%%%%%%%%%%%%%%%%%%%%%%%%%%%%%%%%%%%%%%%%%%

%\title{Galileons from Weyl bi-Connection}
\title{Spontaneous ``Scalar-Vector Galileons''  from ``Weyl bi-Connection'' model}

\author{Nima Khosravi}
\email{nima@ipm.ir}

\affiliation{School of Astronomy, Institute for Research in Fundamental Sciences (IPM), P. O. Box 19395-5531, Tehran, Iran}

\begin{abstract}
Weyl bi-connection model manifests a natural framework to automatically
produce the Galileon structure. It is shown that this framework can explain
scalar Galileon, vector Galileon as well as their interactions by
generalizing the Weyl non-metricity. So it can be interpreted as a
geometrical realization for Galileons. The non-metricity part enjoys a
$U(1)$ gauge invariance. The result is interestingly non-trivial since the
Galileon structure appears spontaneously and not by demanding the absence of
the Ostrogradsky ghost. This fact suggests a possible deeper conceptual
relation between Weyl bi-connection model and the absence of Ostrogradsky
ghost. 

% So maybe Weyleon is a more proper name instead of Galileons!

\end{abstract}
\maketitle
%\tableofcontents
%%%%%%%%%%%%%%%%%%%%%%%%%%%%%%%%%%%%%%%%%%%%%%%%%%%%%%%%%%%%%%%%%%%%%%%%%%%%%%%%%%%%%%%%%%%%
%\newpage
\section{Introduction and Motivations}
To explain the late time acceleration of the universe we need to modify the Einstein-Hilbert general relativity (GR). This can be done by adding the cosmological constant (CC) though this model fits observational data but it suffers from CC-problem. To avoid this problem it is generally believed that dynamical modification of GR may help\footnote{It is not the only reason to seek for modified gravity. The other issue is to explain why the CC density is more or less equal to the present matter energy density. In addition the early universe inflation proposal suggests similar accelerating phase which had an end. So why it cannot happen for late time accelerating phase \cite{euclid}.}. This is not an easy task to do since any modified gravity model should be same as GR at small scales (e.g. solar system) but deviate from GR at the cosmological scales. In addition to this phenomenological property of any successful modified gravity it should also be consistent at the level of theory.  The theoretical problem can be seen when we emphasize modified gravity models usually have to have additional degrees of freedom (dof) \cite{arkani}. The consistency of the model imposes the absence of any ghost in these dof. The simplest modified gravity can be achieved by adding just one scalar field. An interesting candidate for this purpose is Galileon model. The Galileons are the most general ghost free Lagrangians for a scalar field which are non-linear such that can behave same as GR in local scales and differ from GR in large scales. Galileon model has been studied in \cite{nicolis} for a flat background and its covariant version has been reported in \cite{cedric} however it can be showed that it is a subset of Horndeski action \cite{horn}. In the following we review very briefly the Galileon Lagrangians for future purposes.

{\bf Galileon Lagrangians:} Mathematically, absence of the Ostrogradsky ghost in a model  imposes absence of higher than two derivative terms at the level of equations of motion. This means the model should enjoys the Galilean symmetry i.e. $\pi\rightarrow \pi+b^\mu x_\mu+c$ where $\pi$ is our scalar field, $b^\mu$ and $c$ are a constant four vector and a scalar respectively. Demanding Galilean symmetry  at the level of equations of motion  means appearance of $(\partial \partial \pi)^n$ terms in equations of motion. Since $\delta {\cal L}\simeq {\cal E}\delta \pi$ where $\cal E$ is the equation of motion then Galilean symmetry for $\cal E$  means ${\cal L}$ contains terms like $\partial \pi \partial \pi (\partial \partial \pi)^n$. These forms are building blocks for Galileon Lagrangian but the crucial relative coefficients are setting by hand to be sure there is no Ostrogradsky ghost. In four dimensions there are five independent Lagrangians which satisfy the above conditions. They are
\begin{eqnarray}\label{galileon}  \nonumber
{\cal L}^{G.}_1&=&\pi\\\nonumber
{\cal L}^{G.}_2&=&\partial \pi.\partial \pi\\\nonumber
{\cal L}^{G.}_3&=&[\Pi]\partial\pi.\partial\pi\\\nonumber
{\cal L}^{G.}_4&=&[\Pi]^2\partial\pi.\partial\pi-2[\Pi]\partial\pi.\Pi.\partial\pi-[\Pi^2]\partial\pi.\partial\pi+\partial\pi.\Pi^2\partial\pi\\\nonumber
{\cal L}^{G.}_5&=&[\Pi]^3\partial\pi.\partial\pi-3[\Pi]^2\partial\pi.\Pi.\partial\pi-3[\Pi][\Pi^2]\partial\pi.\partial\pi+6[\Pi]\partial\pi.\Pi^2.\partial\pi+2[\Pi^3]\partial\pi.\partial\pi+3[\Pi^2]\partial\pi.\Pi.\partial\pi-6\partial\pi.\Pi^3.\partial\pi
\end{eqnarray}
where $\partial=\partial_\mu$, $[...]$ is taking trace by $g_{\mu\nu}$, $\Pi_{\mu\nu}=\partial_\mu\partial_\nu\pi$ and ``dot'' is Einstein summation rule. The above Lagrangians are unique in four dimensions and result in equations of motion with just two derivatives on $\pi$. As it is obvious from the above results the Lagrangians' structure is exactly $\partial \pi \partial \pi (\partial \partial \pi)^n$ and the relative coefficients are fixed by hand to kill Ostrogradsky ghost. Note that any minor changes in the coefficients may introduce Ostrogradsky ghost so in principle there is a coefficient fine tuning in Galileon terms. After tuning the coefficients by hand then it was found that one can write the above Lagrangian by using the Levi-Civita symbol as
\begin{eqnarray}
{\cal L}^{G.}_{n+1}=\epsilon^{\mu_1\nu_1\mu_2\nu_2...\mu_{2n}\nu_{2n}}\,\partial_{\mu_1}\pi\,\partial_{\nu_1}\pi\,\Pi_{\mu_2\nu_2}\,...\,\Pi_{\mu_n\nu_n}.
\end{eqnarray}
The appearance of the Levi-Civita symbol is crucial in Galileon structure and makes the properties of Galileons very clear and straightforward to understand. This fact plays the same role in vector generalization of Galileon \cite{vector-G}. It is worth to mention that appearance of the Levi-Civita symbol seems to be crucial in the absence of ghosts. In fact the Lovelock gravity is another example. The Lovelock terms are the generalization of Einstein-Hilbert action to higher order derivative terms such that the ghost does not appear. Their structure is as follow
\begin{eqnarray}
{\cal L}^{L.}_{n+1}=\epsilon^{\mu_1\nu_1\mu_2\nu_2...\mu_{2n}\nu_{2n}}\,R_{\mu_1\nu_1\mu_2\nu_2}\,...\,R_{\mu_{2n-1}\nu_{2n-1}\mu_{2n}\nu_{2n}}
\end{eqnarray}
where again appearance of the Levi-Civita symbol is crucial for a healthy model. 

In this paper we are going to show how in Weyl bi-connection model the Levi-Civita symbol appears automatically and not by a prior demanding of the absence of the ghost. Then we show scalar Galileons and vector Galileons (generalized Proca action) can be automatically coming out of Weyl bi-Connection model. For this purpose in the next section we briefly review the Weyl bi-connection model.

\section{Weyl-bi-Connection Model}
In \cite{nima} a model with two connections, $^{(1)}\Gamma^\mu_{\alpha\beta}$ and $^{(2)}\Gamma^\mu_{\alpha\beta}$, has been studied. We assumed the total curvature, ${\cal R}^\mu_{\nu\rho\sigma}$, is a superposition of curvatures corresponds to each connection i.e. $2 {\cal R}^\mu_{\nu\rho\sigma}=R^\mu_{\nu\rho\sigma}(^{(1)}\Gamma^\mu_{\alpha\beta})+R^\mu_{\nu\rho\sigma}(^{(2)}\Gamma^\mu_{\alpha\beta})$ which says the Einstein-Hilbert action is modified to 
\begin{eqnarray}\label{ricci-lag}
{\cal{L}}&=&\sqrt{g}{\cal R}=\sqrt{-g}g^{\mu\nu}\big(R_{\mu\nu}(\gamma)+\Delta_{\mu\nu}\big)
\end{eqnarray}
 where $\gamma^\mu_{\alpha\beta}=\frac{1}{2}\left(^{(1)}\Gamma^\mu_{\alpha\beta}+^{(2)}\Gamma^\mu_{\alpha\beta}\right)$ is the average connection, $\Delta_{\nu\sigma}=\delta^\rho_\mu\Delta^{\mu}_{\nu\rho\sigma}=\delta^\rho_\mu\left(\Omega^\mu_{\alpha\rho} \Omega^\alpha_{\nu\sigma}-\Omega^\mu_{\alpha\sigma} \Omega^\alpha_{\nu\rho}\right)$ and the difference tensor is defined as $\Omega^\mu_{\alpha\beta}=^{(1)}\Gamma^\mu_{\alpha\beta}-^{(2)}\Gamma^\mu_{\alpha\beta}$. According to Palatini approach the equation of motion with respect to $\gamma^\mu_{\alpha\beta}$ imposes $\gamma^\mu_{\alpha\beta}$ to be the Christoffel symbol corresponding to metric $g_{\mu\nu}$. So the above Lagrangian can be written as
\begin{eqnarray}\label{lag}
{\cal{L}}&=&\sqrt{-g}\big(R+\Delta\big)
\end{eqnarray}  
where $R$ is Ricci scalar of the metric $g_{\mu\nu}$ and $\Delta=g^{\nu\sigma}\Delta_{\nu\sigma}$. As we showed in \cite{nima} an interesting case in bi-connection model can be inspired by Weyl geometry. In Weyl geometry the connection does not satisfy the metricity. The non-metricity is modeled by introducing a vector such that $\nabla_\mu g_{\alpha\beta}=C_\mu g_{\alpha\beta}$. Physically, existence of this vector says parallel transportation along a geodesic not only changes the direction of a given vector but also its amplitude.  An interesting example of Weyl geometry is for $C_\mu=\partial_\mu \phi$ where $\phi$ is a scalar field. For this case the connection (i.e. covariant derivative) corresponds to a metric, $\tilde g_{\mu\nu}$, such that $\tilde g_{\mu\nu}=e^{-\phi}g_{\mu\nu}$. In this work we will focus on this interesting case of Weyl geometry in bi-connection framework. But before that let us study a generalization of Weyl geometry in bi-connection context.

Inspired by Weyl geometry it is assumed that the connections satisfy the following relations $^{(1)}\nabla_\mu
g_{\alpha\beta}=-C_\mu X_{\alpha\beta}$ and $^{(2)}\nabla_\mu
g_{\alpha\beta}=+C_\mu X_{\alpha\beta}$ where $^{(1)}\nabla_\mu$ and
$^{(2)}\nabla_\mu$ are covariant derivatives with respect to
$^{(1)}\Gamma^\alpha_{\mu\nu}$ and $^{(2)}\Gamma^\alpha_{\mu\nu}$
respectively and $X_{\alpha\beta}$ is an arbitrary symmetric tensor. It is straightforward to show that
\begin{eqnarray}\label{2-connection-WG-higher-order}
^{(1)}\Gamma^\alpha_{\mu\nu}&=&\big\{^\alpha_{\mu\nu}\big\}+\frac{1}{2}g^{\alpha\beta}(
X_{\nu\beta}C_\mu+ X_{\mu\beta}C_\nu-
X_{\mu\nu}C_\beta),\\\nonumber
^{(2)}\Gamma^\alpha_{\mu\nu}&=&\big\{^\alpha_{\mu\nu}\big\}-\frac{1}{2}g^{\alpha\beta}(
X_{\nu\beta}C_\mu+ X_{\mu\beta}C_\nu-
X_{\mu\nu}C_\beta)
\end{eqnarray}
where $\big\{^\alpha_{\mu\nu}\big\}$ is the Christoffel symbol corresponding to $g_{\mu\nu}$. It is also obvious that the average connection is $\gamma^\alpha_{\mu\nu}=\big\{^\alpha_{\mu\nu}\big\}$ in agreement with Palatini approach and the difference tensor is $\Omega^\alpha_{\mu\nu}=g^{\alpha\beta}(
X_{\nu\beta}C_\mu+ X_{\mu\beta}C_\nu-X_{\mu\nu}C_\beta)$. So in the Lagrangian (\ref{lag}), $\Delta$ will be
\begin{eqnarray}\label{weyl-bi-lag}
\Delta=2C^\mu C^\nu \big(X X_{\mu\nu}-X_{\mu\rho}X^\rho_\nu\big)+C^2\big(X_{\mu\nu}X^{\mu\nu}-X^2\big).
\end{eqnarray}
Interestingly, the above Lagrangian can be written as 
\begin{eqnarray}\label{weyl-bi-lag2}
\Delta=-\epsilon^{\alpha\mu\rho}_{\beta\nu\sigma}\,C_\alpha \,C^\beta\,X_\mu^\nu\,X_\rho^\sigma
\end{eqnarray}
where $\epsilon^{\alpha\mu\rho}_{\beta\nu\sigma}$ is Levi-Civita symbol. This property is non-trivial and a pivotal fact about Weyl bi-connection model. It should be emphasized that the Levi-Civita symbol appears automatically and it is not because of any prior fine tuning of coefficients. So Weyl bi-connection model has this very natural property which makes the formalism a natural framework of all kind of Galileons including scalar, multi-scalar, vector, scalar-vector and massive gravity.

\subsection{Scalar Galileons in Weyl-bi-Connection model:}
Now let us back to the interesting example of conformal Weyl geometry. This case is equivalent to $C_\mu=\partial_\mu \pi$ where $\pi$ is a scalar field and $X_{\mu\nu}=g_{\mu\nu}$ which results in $\Delta=-6 \partial_\mu \pi \partial^\mu \pi$ which is the standard kinetic term for a scalar field. But there is a very significant generalization of this case by assuming
\begin{eqnarray}\label{x-galileon}
X_{\mu\nu}=g_{\mu\nu}+\alpha \,  \Pi_{\mu\nu} \hspace{1cm}\equiv\hspace{1cm} \nabla_\alpha g_{\mu\nu}=\partial_\alpha \pi \big(g_{\mu\nu}+\alpha \, \Pi_{\mu\nu}\big)
\end{eqnarray}
 where $\Pi_{\mu\nu}=\partial_\mu \partial_\nu \pi$ and $\alpha$ is an arbitrary  constant. With this generalization we get
\begin{eqnarray}\label{bi-con-galileon-lag}
\Delta_{G.}=-6 \partial \pi.\partial \pi-4\alpha\bigg(\partial\pi.\partial\pi [\Pi]-\partial\pi.\Pi.\partial\pi\bigg)-\alpha^2\bigg(\partial\pi.\partial\pi[\Pi]^2-\partial\pi.\partial\pi[\Pi^2]+2\partial\pi.\Pi^2.\partial\pi-2[\Pi]\partial\pi.\Pi.\partial\pi\bigg)
\end{eqnarray}
 The above Lagrangian\footnote{Note that we have this relation: $\partial\pi.\partial\pi [\Pi]-\partial\pi.\Pi.\partial\pi\doteq \frac{3}{2}\partial\pi.\partial\pi [\Pi]$ where $\doteq$ is equality up to a total derivative.} is a surprise result since the above terms are what are well-known as Galileon terms excluding the ${\cal L}^{G.}_5$ in four dimensions. We should emphasize that the model automatically produces all the terms with correct coefficients without any prior fine-tuned postulate.

\subsection{Vector Galileons (or generalized Proca) in Weyl-bi-Connection model:}
Now let us assume $C_\mu=A_\mu$ and
\begin{eqnarray}\label{x-vector-galileon}
X_{\mu\nu}=g_{\mu\nu}+\alpha \,  (\partial_\mu A_\nu+\partial_\nu A_\mu) \hspace{1cm}\equiv\hspace{1cm} \nabla_\alpha g_{\mu\nu}=A_\alpha \big(g_{\mu\nu}+2\alpha \, \partial_{(\mu}A_{\nu)}\big),
\end{eqnarray}
then the Lagrangian (\ref{weyl-bi-lag}) will be
\begin{eqnarray}\label{bi-con-vector-galileon-lag}
\Delta_{V.G.}&=&-6 A.A-4\alpha\bigg(A.A\, \partial.A-A^\mu\,A^\nu\,\partial_\mu A_\nu\bigg)\\\nonumber&-&\alpha^2\bigg(A.A\,(\partial.A)^2-A.A\,\partial_\mu A_\nu \, \partial^\mu A^\nu+2A_\mu\,A_\nu\,\partial ^\mu A_\rho\,\partial^\rho A^\nu-2\partial.A \,A^\mu\,A^\nu\,\partial_\mu A_\nu\bigg)
\end{eqnarray}
which is exactly same as the Lagrangians in \cite{vector-G} where the coefficients are tuned by hand. 

\subsection{Scalar-Vector Galileons  in Weyl-bi-Connection model:}
It is also interesting to study the combination of a scalar and a vector in this formalism. To do that we assume $C_\mu=\partial_\mu \pi+A_\mu$ and
\begin{eqnarray}\label{x-vector-galileon}
X_{\mu\nu}=g_{\mu\nu}+\alpha \,\left[\Pi_{\mu\nu}+  (\partial_\mu A_\nu+\partial_\nu A_\mu)\right] \hspace{1cm}\equiv\hspace{1cm} \nabla_\alpha g_{\mu\nu}=(\partial_\alpha \pi+A_\alpha) \big(g_{\mu\nu}+\alpha \Pi_{\mu\nu}+2\alpha \, \partial_{(\mu}A_{\nu)}\big),
\end{eqnarray}
then the mixing scalar-vector part of the Lagrangian (\ref{weyl-bi-lag}) will be
\begin{eqnarray}\label{bi-con-scalar-vector-galileon-lag}
\Delta_{S.V.G.}&=&-12A.\partial \pi\\\nonumber&-&4 \alpha\bigg[\partial\pi.\partial\pi\,\partial.A+A.A[\Pi]+2\partial.A\,A.\partial\pi+[\Pi]A.\partial\pi-\partial^\mu \pi
\,\partial^\nu\,\partial_\mu A_\nu-A^\mu A^\nu\Pi_{\mu\nu}\\\nonumber&-&A^\nu\,\partial_\mu A_\nu\,\partial^\mu\pi-A^\mu \partial^\nu\pi\,\partial_\nu A_\nu-\Pi_{\mu\nu}\,\partial^\mu\pi\,A^\nu-\Pi_{\mu\nu}\,\partial^\nu\pi\,A^\mu\bigg]+{\cal O}(\alpha^2)
\end{eqnarray}
and the ${\cal O}(\alpha)$ term can be written as (up to a total derivative)
\begin{eqnarray}\label{bi-con-scalar-vector-galileon-lag}
\Delta^{(\alpha)}_{S.V.G.}&=&-6\alpha\bigg[\partial\pi.\partial\pi\,\partial.A+2 \Box\pi\, A.\partial\pi+A.A\,\Box\pi+2 \partial.A\,A.\partial\pi\bigg]
\end{eqnarray}
where the first two terms are exactly what is found in \cite{vector-G}. But there are two additional term in this model which are also free of Ostrogradsky ghost. Note that the first two terms are second order in $\pi$ and first order in $A$ but the last two terms have the opposite situation.

To see if the behaviour of ${\cal O}(\alpha^2)$ is healthy or not let us recall the Levi-Civita symbol. The Lagrangian will be as follow
\begin{eqnarray}\label{weyl-scalar-vector}
\Delta=-\epsilon^{\alpha\mu\rho}_{\beta\nu\sigma}\,\bigg(\partial_\alpha\pi+A_\alpha\bigg) \,\bigg(\partial^\beta\pi+A^\beta\bigg)\,\bigg(g_{\mu}^{\nu}+\alpha \,\left[\Pi_\mu^\nu+  (\partial_\mu A^\nu+\partial^\nu A_\mu)\right]\bigg)\bigg(g_{\rho}^{\sigma}+\alpha \,\left[\Pi_\rho^\sigma+  (\partial_\rho A^\sigma+\partial^\sigma A_\rho)\right]\bigg)
\end{eqnarray}
and then
\begin{eqnarray}\label{weyl-scalar-vector-alpha2}
\Delta^{(\alpha^2)}=-\epsilon^{\mu_1\nu_1\mu_2\nu_2\mu_3\nu_3}\,\bigg(\partial_{\mu_1}\pi+A_{\mu_1}\bigg) \,\bigg(\partial_{\nu_1}\pi+A_{\nu_1}\bigg)\,\bigg(\Pi_{\mu_2\nu_2}+  (\partial_{\mu_2} A_{\nu_2}+\partial_{\nu_2} A_{\mu_2})\bigg)\bigg(\Pi_{\mu_3\nu_3}+  (\partial_{\mu_3} A_{\nu_3}+\partial_{\nu_3} A_{\mu_3})\bigg)
\end{eqnarray}
which is easy to show that corresponding equations of motion have just up to second order derivatives thanks to the Levi-Civita symbol.

 \begin{comment} The equations of motion with respect to $A$ is as follow
\begin{eqnarray}\label{eq-vector}
{\cal E}_{V}&=&\varepsilon^{\mu_1\nu_1\mu_2\nu_2\mu_3\nu_3}\,\delta_{\mu_1}^\mu \,\bigg(\partial_{\nu_1}\pi+A_{\nu_1}\bigg)\,\bigg(\Pi_{\mu_2\nu_2}+  (\partial_{\mu_2} A_{\nu_2}+\partial_{\nu_2} A_{\mu_2})\bigg)\bigg(\Pi_{\mu_3\nu_3}+  (\partial_{\mu_3} A_{\nu_3}+\partial_{\nu_3} A_{\mu_3})\bigg)\\\nonumber
&+&\varepsilon^{\mu_1\nu_1\mu_2\nu_2\mu_3\nu_3}\,\delta_{\nu_1}^\mu \,\bigg(\partial_{\mu_1}\pi+A_{\mu_1}\bigg)\,\bigg(\Pi_{\mu_2\nu_2}+  (\partial_{\mu_2} A_{\nu_2}+\partial_{\nu_2} A_{\mu_2})\bigg)\bigg(\Pi_{\mu_3\nu_3}+  (\partial_{\mu_3} A_{\nu_3}+\partial_{\nu_3} A_{\mu_3})\bigg)
\end{eqnarray}

\end{comment}

\subsection{Multi Scalar-Vector Galileons  in Weyl-bi-Connection model:}
In a generalization of above examples one can assume $C_\mu=\Sigma_I\partial_\mu \pi^I+\Sigma_J A^J_\mu$ and $X_{\mu\nu}=g_{\mu\nu}+\,\left[\Sigma_I \alpha^I \Pi^I_{\mu\nu}+\Sigma_J  \alpha^J (\partial_\mu A^J_\nu+\partial_\nu A^J_\mu)\right]$ for a multi scalar-vector case. This form of multi scalar-vector Galileon is ghost free since one can think of $\Sigma_I \pi$ and $\Sigma_J A_\mu$ as a scalar and vector field respectively and then reduce this case to the previous one.

\section{Questions and Discussions}
It seems Weyl bi-Connection model is a natural framework for working on Galileons. But there are some related issues at this level
\begin{itemize} 
\item  {\bf Covariantization:} The above Galileon terms are for flat spacetime and one needs to generalize them for curved background (i.e. covariant Galileons).
\item {\bf Fifth Galileon term:} The other important issue is how one can get ${\cal L}^{G.}_5$ in this formalism?
\end{itemize}

To have a clue for above questions a way is to think on generalization of $X_{\mu\nu}$ tensor. In the following sub-section we try to give some properties of mathematical structure of this framework as  possible hints to address the above questions.

\subsection{Some mathematical properties}
Let us generalize (\ref{x-galileon}) by assuming $X_{\mu\nu}=g_{\mu\nu}+\alpha \,  \Pi_{\mu\nu}+\beta \, \Phi_{\mu\nu}$ where $\Phi_{\mu\nu}$ is an arbitrary symmetric tensor. Because of this additional term we will have additional term in (\ref{bi-con-galileon-lag}) such that
\begin{eqnarray}\label{generalized-lag}
\Delta=\Delta_{G.}&+&4\beta\bigg(\partial\pi.\Phi.\partial\pi-\partial\pi.\partial\pi [\Phi]\bigg)\\\nonumber&+&2\alpha\beta\bigg(\partial\pi.\partial\pi[\Pi\Phi]-\partial\pi.\partial\pi[\Pi][\Phi]+[\Pi]\partial\pi.\Phi.\partial\pi+[\Phi]\partial\pi.\Pi.\partial\pi-2\partial\pi.\Pi.\Phi.\partial\pi\bigg)\\\nonumber&+&\beta^2\bigg(\partial\pi.\partial\pi[\Phi^2]-\partial\pi.\partial\pi[\Phi]^2+2[\Phi]\partial\pi.\Phi.\partial\pi-2\partial\pi.\Phi^2.\partial\pi\bigg).
\end{eqnarray}
Now let us check some specific forms of $\Phi_{\mu\nu}$:\\
\itemize{
\item {\bf $\Phi_{\mu\nu}=\partial_\mu\pi\partial_\nu \pi$}: In this case interestingly the additional terms in (\ref{generalized-lag}) vanish.

\item {\bf $\Phi_{\mu\nu}=a\, G_{\mu\nu}+b\,R\,g_{\mu\nu}$}: This case may have a hint for covariantization of the above model as we show. The idea in \cite{cedric} is to add derivatives of metric to compensate the commutation of indices on $\partial_\mu\partial_\nu\pi$. Here we need to check if there is a possible solution for $a$ and $b$ such that the correct term for covariantization of ${\cal L}^{G.}_4$ can be appeared. The first additional term in (\ref{generalized-lag}) will be
\begin{eqnarray}
4\beta\bigg(\partial\pi.\Phi.\partial\pi-\partial\pi.\partial\pi [\Phi]\bigg)=4\beta\bigg(a\,G^{\mu\nu}\,\partial_\mu\pi\partial_\nu\pi -(3b-a)\,R\,\partial_\mu\pi\partial^\mu\pi \bigg)
\end{eqnarray}
where for the case ``$3b=a$'' and ``$4\beta\,a=-\alpha^2\,\partial_\pi\pi\partial^\mu\pi$'' the result is exactly what we need for the fourth covarinat Galileon \cite{cedric}. But there is a remaining issue: the second and third lines in (\ref{generalized-lag}) also produce some terms. However these terms are higher order terms and their considerations may need considering higher order gravity from the beginning e.g. adding Lovelock terms to (\ref{ricci-lag}). We remain this maybe interesting problem for future works. 

\item {\bf $\Phi_{\mu\nu}=a\,\Pi_{\mu\rho}\,\Pi^\rho_\nu+b\,[\Pi]\,\Pi_{\mu\nu}$}: This case can have a hint for producing the fifth term of Galileon. On the other hand its structure reminds us the decoupling limit procedure (though not exactly because of $b$-term) which is used to decouple scalar mode and tensor mode e.g. in massive gravity analysis \cite{arkani2}. The first additional term (i.e. the first line in (\ref{generalized-lag})) will be as follow 
\begin{eqnarray}
4\beta\bigg(b\,\partial\pi.\partial\pi[\Pi]^2-a\,\partial\pi.\partial\pi[\Pi^2]+a\,\partial\pi.\Pi^2.\partial\pi-b\,[\Pi]\partial\pi.\Pi.\partial\pi\bigg)
\end{eqnarray}
which has common terms with ${\cal L}^{G.}_4$ but with incorrect coefficients. So there is a danger of ghost appearance. But in flat spacetime there is a relation which says $\partial\pi.\partial\pi[\Pi]^2-\partial\pi.\partial\pi[\Pi^2]-2\partial\pi.\Pi^2.\partial\pi+2[\Pi]\partial\pi.\Pi.\partial\pi$ is a total derivative. This means we are secured if we assume ``$a=-b$'' in the above additional term. By this assumption this additional term is proportional to ${\cal L}^{G.}_4$. Now it is time for the second line terms in (\ref{generalized-lag}) which is proportional to:
\begin{eqnarray}\nonumber
&&\bigg([\Pi]^3\partial\pi.\partial\pi-2[\Pi]^2\partial\pi.\Pi.\partial\pi-2[\Pi][\Pi^2]\partial\pi.\partial\pi+3[\Pi]\partial\pi.\Pi^2.\partial\pi+[\Pi^3]\partial\pi.\partial\pi+[\Pi^2]\partial\pi.\Pi.\partial\pi-2\partial\pi.\Pi^3.\partial\pi\bigg)\\\nonumber
&&={\cal L}^{G.}_5+\bigg([\Pi]^2\partial\pi.\Pi.\partial\pi+[\Pi][\Pi^2]\partial\pi.\partial\pi-3[\Pi]\partial\pi.\Pi^2.\partial\pi-[\Pi^3]\partial\pi.\partial\pi-2[\Pi^2]\partial\pi.\Pi.\partial\pi+4\partial\pi.\Pi^3.\partial\pi\bigg)
\end{eqnarray}
which differs from ${\cal L}^{G.}_5$ but at least it has the same terms with inappropriate coefficients. One may think that the higher order Lovelock term can resolve this issue. Note that in this case we also have an additional term coming from the third line of (\ref{generalized-lag}) which should be considered.\\
}

Though in above we gave some ideas to address the covariant Galileon as well as the fifth Galileon term but it seems for a concrete answer we need more considerations.

{\bf $U(1)$ non-metricity:} It seems there is another interesting property of the above formalism. To see that let us remind $C_\mu$ and $X_{\mu\nu}$ for example in case of scalar-vector
$C_\mu=\partial_\mu \pi+A_\mu$ and $X_{\mu\nu}=g_{\mu\nu}+\alpha \,\left[\Pi_{\mu\nu}+  (\partial_\mu A_\nu+\partial_\nu A_\mu)\right]$ and the corresponding non-metricity relation
\begin{eqnarray}\label{x-vector-galileon1}
\nabla_\alpha g_{\mu\nu}=(\partial_\alpha \pi+A_\alpha) \big(g_{\mu\nu}+\alpha \Pi_{\mu\nu}+2\alpha \, \partial_{(\mu}A_{\nu)}\big).
\end{eqnarray}
By looking more carefully at the above relation we can see $X_{\mu\nu}-g_{\mu\nu}=2\alpha\partial_{(\mu}C_{\nu)}$. So for a given $C_\mu$, $X_{\mu\nu}$ is totally fixed.  In addition the structure of Weyl bi-connection model reminds us a $U(1)$ gauge symmetry for the non-metricity part of the model i.e. $C_\mu$ and $X_{\mu\nu}$ as 
\begin{eqnarray}\label{u1}
A_\mu\rightarrow A_\mu+\partial_\mu\,\Lambda\,\,\,\,  and\,\,\,\, \pi\rightarrow \pi-\Lambda  \,\,\Longrightarrow\,\, C_\mu\rightarrow C_\mu\,\,\,\, and\,\,\,\, X_{\mu\nu}\rightarrow X_{\mu\nu}
\end{eqnarray}
for an arbitrary scalar $\Lambda$.\\

\section{Conclusions: Galileon or Weyleon?}
It has been shown that a ``Weyl geometrical inspired bi-connection formulation of geometry'' provides a natural framework for Galileon structures. The key point in this model is that we do not need to assume ghost-freeness of the model as an extra condition on the model. The model automatically is ghost-free due to automatic appearance of the Levi-Civita symbol in this model. To make this issue clear let us briefly remind what happens for example in Galileon theory. We demand for the most general equations of motion for a scalar field which does not have any extra (ghost) degrees of freedom. This says the Lagrangian should have terms like ``$\partial \pi \,\partial\pi\, \partial\partial \pi$'' but it is not sufficient and we need to fix the relative coefficients to make the model ghost-free. This last process is done by hand.  However in Weyl bi-Connection model the appearance of the Levi-Civita symbol is automatic which results in a natural framework for producing Galileon Lagrangians including scalar,  vector and scalar-vector models. At this level an interesting question is if massive gravity can be a special example of this model. Also
\item {\bf Fundamental symmetry:} Is the above result just an accident or is there  some deeper fact behind it e.g. a kind of symmetry? I.e. what is behind combination of the Weyl geometry and bi-connection model which makes Galileon's appearance spontaneously?\\

To summarize: Weyl-bi-connection model is  a geometrical realization of Galileon. In addition in this formalism, the ghost is automatically killed hence maybe it is more relevant to name the Galileons, Weyleons!

%\newpage

\begin{acknowledgments}
We are grateful to S. Baghram for useful discussions. We also thank K. Koyama, H. R. Sepangi and G. Tasinato for their comments.
\end{acknowledgments}

%===============================================================================
%=============================================================
%\newpage

%=============================================================

\end{document}